\documentclass[aps,prl,twocolumn,superscriptaddress]{revtex4-2} 
\usepackage{amsmath,amssymb,graphicx,bm}
\usepackage{subcaption}
\usepackage{multirow}
\usepackage[colorlinks=true,linkcolor=blue,citecolor=blue,urlcolor=blue]{hyperref}

\begin{document}

\title{Revisiting the Topological Nature of TaIrTe$_4$, SrSi$_2$, and \\Cu$_2$XY$_3$: An \textit{ab-initio} Investigation}

\author{Prakash Pandey}
\altaffiliation{prakashpandey6215@gmail.com}
\affiliation{School of Physical Sciences, Indian Institute of Technology Mandi, Kamand - 175075, India}
\author{Sudhir K. Pandey}
\altaffiliation{sudhir@iitmandi.ac.in}
\affiliation{School of Mechanical and Materials Engineering, Indian Institute of Technology Mandi, Kamand - 175075, India}

\date{\today}

\begin{abstract}

Several topological electronic materials have been theoretically predicted, leading to a comprehensive catalog systematically characterized by their band crossings. Researchers have attempted to experimentally verify the topological nature of some materials from the present catalogs, but not all efforts have yielded positive results.
Here, we introduce a possible reason for the discrepancies between theoretical and experimental results.
In this direction, firstly we have revisited the nature of the well-known topological materials TaIrTe$_4$ and SrSi$_2$ using \textit{state-of-the-art ab-initio} calculations, and found additional Weyl points in both materials that were missing in previously reported studies.
Then we have verified the recently predicted topological states of the \textit{Imm2}-phase of Cu$_2$XY$_3$ (X=Si, Ge, Sn \& Y=S, Se, Te).
Contrary to previously reported results, we did not find any Weyl points or nodal arcs in Cu$_2$SnTe$_3$. Notably, our theoretical results reveal that Cu$_2$SiTe$_3$, Cu$_2$GeTe$_3$ and Cu$_2$GeSe$_3$ each host four small nodal rings, eight Weyl points, and eight nodal arcs, respectively, which differ from previous studies. 
Considering Cu$_2$SnS$_3$ as an example, we have also investigated the robustness of the topological phase against local strain.
Our study provides insights into the inconsistencies between theoretical predictions and experimental results, and demonstrates how the topological phase is sensitive to changes in lattice parameters, atomic positions, and exchange-correlation functionals. 

\end{abstract}

\maketitle

{\it Introduction.|}
In recent years, many topological electronic materials have been predicted, and efforts have been made to systematically catalog them based on the number and type of their band crossings~\cite{zhang2019catalogue, bradlyn2017topological, doi:10.1126/science.aaf5037, PhysRevB.101.245117}. 
Among these, researchers have attempted to experimentally verify the topological nature of some materials and have found that not all of them yield positive results~\cite{PhysRevLett.115.176404, PhysRevB.92.075107, 10.1073/pnas.1601262113, PhysRevLett.116.176803, PhysRevX.6.021017, PhysRevLett.117.237601, HARTSTEIN2020101632, PhysRevResearch.6.033195, PhysRevB.104.075131, PhysRevB.93.201101, PhysRevB.95.241108, belopolski2017signatures, PhysRevB.97.241102, PhysRevB.95.241108, zeng2015topological, PhysRevLett.117.127204,  PhysRevLett.117.127204, PhysRevB.94.165163, PhysRevB.96.041120, PhysRevB.93.235142, dey2018bulk, PhysRevB.92.115119, 10.1126/science.aan4596, zhang2019multiple, borisenko2020strongly, PhysRevX.14.021043}.
With the rapidly growing interest in topological physics, it is important to ask: What might be the possible reasons behind the discrepancies between experimental and theoretical results?
Surprisingly, the massive hunt for new topological materials is still ongoing~\cite{gibney2018thousands, doi:10.1126/sciadv.abd1076, 10.1002/adma.202204113, regnault2022catalogue, rasul2024machine}, rather than bridging the gap between prediction and reality.
Thus, we attempt to address this question by examining key examples of topological materials.

Starting with the example of TaIrTe$_4$, a topological semimetal predicted to be a type-II Weyl semimetal, let us explore the discrepancies between theoretical predictions and experimental observations.
Using maximally localized Wannier functions (MLWFs)~\cite{RevModPhys.96.045008, RevModPhys.84.1419, WU2017}, Koepernik \textit{et al.}~\cite{PhysRevB.93.201101} have predicted that TaIrTe$_4$ is a Weyl semimetal that hosts only four Weyl points. 
Haubold \textit{et al.}~\cite{PhysRevB.95.241108} verified the presence of four Weyl points using angle-resolved photoemission spectroscopy (ARPES).
Interestingly, a later experimental investigation using ARPES~\cite{PhysRevB.97.241102} reported the presence of two nodal lines and twelve Weyl points~\cite{PhysRevB.95.241108}.
Surprisingly, in both experimental studies~\cite{PhysRevB.95.241108, PhysRevB.97.241102}, only a patch of intensity appears near the Fermi level. From this intensity alone, it is difficult to determine how many Weyl nodes or nodal lines are present in the system, highlighting a limitation of the ARPES technique. In such cases, ARPES is handicapped without theoretical support to identify the topological features. More importantly, it appears that the reported experimental results~\cite{PhysRevB.95.241108, PhysRevB.97.241102} are completely driven by theoretical predictions. Therefore, greater emphasis should be placed on the accuracy of theoretical results rather than relying solely on experimental data. 
Another prominent example is SrSi$_2$, which also remains a topic of debate regarding the authenticity of its topological nature. Some experimental studies report a gapless semimetallic nature~\cite{10.1063/1.4772973, 10.3389/fchem.2014.00106}, while others suggest it is a narrow-gap semiconductor based on transport measurements~\cite{PhysRevB.110.224514}. Interestingly, some theoretical studies using hybrid functionals have also reported that the Weyl points vanish and a band gap of up to 80 meV opens~\cite{PhysRevB.110.224514}, whereas another theoretical study claims that the topological nodes persist even with hybrid functionals~\cite{singh2018tunable}.

Considering the above examples, it is important to understand the possible reasons for the discrepancies between experimental and theoretical results, as well as between different theoretical results. The discrepancies between experimental and theoretical results arise due to several reasons. Firstly, theoretical calculations are based on idealized models, such as perfect crystalline structures, absence of impurities, and ideal boundary conditions, all of which are approximations. However, real materials inherently contain defects, disorder, and interactions. Secondly, techniques like ARPES are intended to serve as benchmark experimental methods for identifying the Dirac/Weyl phase of a material, but they also require theoretical support. Therefore, the choice of measurement technique may also lead to different results for the same material~\cite{shao2019optical, mele2019dowsing}. 

The discrepancies between different theoretical results may arise from differences between results obtained directly from \textit{first-principles} methods, such as density functional theory (DFT), and those derived from DFT calculations through Wannierization. It is well known that each DFT code uses a different basis set (such as plane waves, localized atomic orbitals or augmented plane waves), which affect the accuracy and convergence of the computed material properties. Due to this, one should converge with respect to calculation parameters such as cutoff energy, \textit{k}-mesh sampling, number of bands, and smearing width to ensure that the results are reliable and physically meaningful. Since predicting topological properties using the DFT method is more computationally expensive than using the tight-binding model (TBM) with MLWFs~\cite{RevModPhys.96.045008, RevModPhys.84.1419, WU2017}. Moreover, the TBM is highly sensitive to various parameters, such as the disentangled energy window, the weight of the projectors, the number \& type of Wannier functions, and the \textit{k}-mesh sampling~\cite{PANDEY2023108570, Pandey_2021}. Therefore, the topological properties should converge with respect to these parameters, and the results obtained through DFT and Wannierization should be exactly the same. Suppose a situation in which the topological properties of materials do not properly converge with respect to the calculation parameters at DFT level, and the properties at the Wannierization level are also not converged; this contributes to the overall inaccuracy. 
Therefore, in such circumstances, the reliability of predicting a material's topological features through Wannierization is questionable.
Recently, it has been observed that the \texttt{PY-NODE} code, based on a \textit{first-principles} approach, provides better results for various materials~\cite{PANDEY2023108570, pandey2025realization, Pandey_2023} than the TBM, making it a convincing tool for benchmarking the materials.

Motivated by the inconsistencies between the predicted theoretical results for TaIrTe$_4$ and SrSi$_2$, it is necessary to revisit and verify the topological nature of the Cu$_2$XY$_3$ (X=Si, Ge, Sn \& Y=S, Se, Te) class of materials using a \textit{first-principles} approach. Zhou \textit{et al.}~\cite{PhysRevResearch.4.033067} recently predicted the topological nature of Cu$_2$XY$_3$ class of materials using a TBM based on MLWF. However, due to limitations in the Wannierization procedure for obtaining reliable TBM, there may be a high possibility that topological insulators may be misidentified as topological semimetals, and vice versa. Furthermore, these materials have not yet been explored experimentally from a topological perspective. Therefore, it is necessary to verify the topological nature of each material belonging to the Cu$_2$XY$_3$ class.

In this work, based on \textit{first-principles} calculations, firstly we have revisited the nature of the well-known topological materials TaIrTe$_4$ and SrSi$_2$, and report extra Weyl points in both materials that were lacking in previous studies. 
Then we have verified the recently predicted topological states of the Cu$_2$XY$_3$ (X=Si, Ge, Sn \& Y=S, Se, Te)~\cite{PhysRevResearch.4.033067} class of materials and compared them with previously reported results. Contrary to previously reported results~\cite{PhysRevResearch.4.033067}, our theoretical findings reveal that Cu$_2$SnTe$_3$, like Cu$_2$SnSe$_3$, does not exhibit any Weyl points or nodal rings. Moreover, our theoretical results reveal that Cu$_2$SiTe$_3$ (Cu$_2$GeTe$_3$) hosts four small nodal rings (eight Weyl points) only, which differs from previous results~\cite{PhysRevResearch.4.033067}, highlighting the inaccuracy of the constructed TBM.
Furthermore, using Cu$_2$SnS$_3$ as an example, we have examined the robustness of topological phases against atomic positions (i.e., local strain), which has been completely untouched from a topological perspective.

{\it Computational details.|}
The electronic structure calculations within DFT theory are performed using \texttt{WIEN2k} package~\cite{blaha2020wien2k}. 
The generalized gradient approximation (GGA)~\cite{PhysRevLett.77.3865} within the Perdew-Burke-Ernzerhof (PBE) is used as the exchange-correlation (XC) functional for the self-consistent energy calculations of the materials TaIrTe$_4$, SrSi$_2$, and Cu$_2$XY$_3$. 
The calculations use the optimized lattice constants and Wyckoff positions for all the materials belonging to the Cu$_2$SnS$_3$ family (see Table I in the Supplementary Material (SM)~\cite{spl}). 
In order to calculate the ground state energy of TaIrTe$_4$, SrSi$_2$, and Cu$_2$XY$_3$, 22$\times$6$\times$6, 16$\times$16$\times$16, and 12$\times$12$\times$12 $k$-mesh sizes are used, respectively, with the energy convergence limit of $10^{-8}$ Ry/cell. 
To calculate the topological properties such as Weyl/line node, we have used \texttt{PY-NODE} code~\cite{PANDEY2023108570}. This code uses Nelder-Mead's function minimization method~\cite{10.1093/comjnl/7.4.308, PANDEY2024109281} to search the node point. To calculate the chirality for each Weyl point, we have used the \texttt{WloopPHI} code~\cite{saini2022wloopphi}.

{\it Topological nature of TaIrTe$_4$.|}
TaIrTe$_4$ belongs to a noncentrosymmetric orthorhombic structure with space group Pmn2$_1$ (No. 31). The lattice constants of this material are $a = 3.795$ \AA, $b = 12.474$ \AA, and $c = 13.242$ \AA. Haubold \textit{et al.}~\cite{PhysRevB.95.241108} predicted and experimentally identified TaIrTe$_4$ as a type-II Weyl semimetal hosting only four Weyl points.
However, later combined experimental and theoretical studies~\cite{PhysRevB.97.241102} suggest that this material exhibits twelve Weyl points and two nodal lines. 
Motivated by the discrepancies between earlier reported results, we performed electronic structure calculations using a \textit{first-principles} DFT-based code to verify the previously reported results. 
The corresponding band structure in the presence of SOC is shown in Fig. \ref{Fig.bands_TaIrTe}.
\texttt{PY-NODE} code successfully identified the touching point between the topmost valence band and the bottommost conduction band across the full Brillouin zone. 
We found a total of sixteen Weyl nodes and two nodal lines. Notably, four additional Weyl points found in our analysis were absent in previous studies~\cite{PhysRevB.97.241102}.
The coordinates and corresponding energies of all the Weyl points we found are listed in Table \ref{table_TaIrTe}.
To check whether these four extra node points are Weyl points or not, we calculated their chiralities ($\chi$) and found that their chiralities are $\chi \in \left\lbrace -1, 1\right\rbrace $.  
Our results highlight that the tight-binding model constructed using MLWFs in previous studies was inaccurate. Therefore, it should be ensured that the level of accuracy is the same as DFT. In this regard, the \texttt{PY-NODE} code is better equipped to identify all the nodal points present in the material.


\begin{figure}\label{Fig.bands_TaIrTe}
\includegraphics[width=0.90\linewidth, height=5cm]{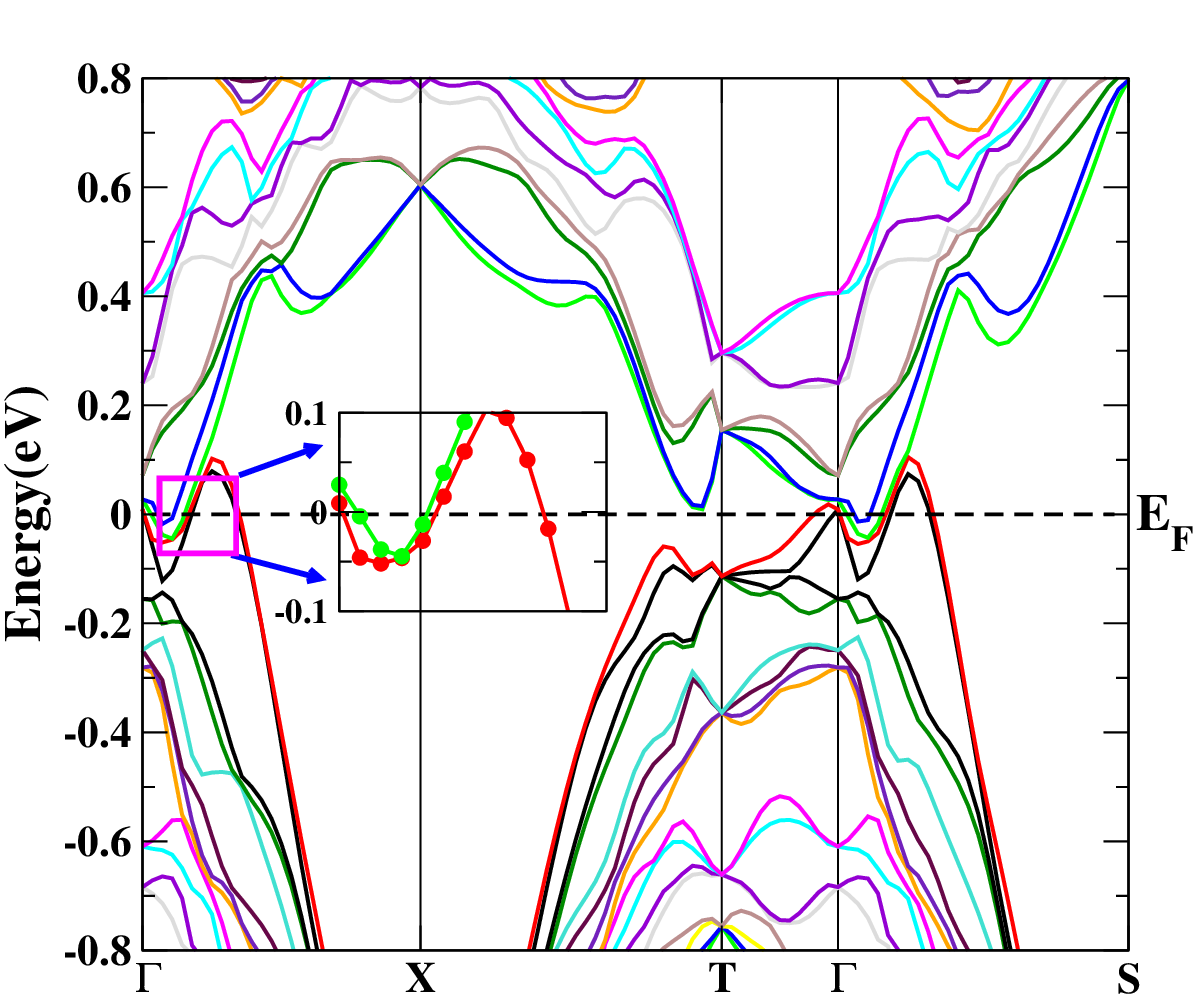}
\caption{\label{Fig.bands_TaIrTe}\small{(Color online)
Calculated bulk band structure of TaIrTe$_4$ in the presence of spin-orbit coupling. Zero energy represents the Fermi level (E$_F$).
}}
\end{figure}

\begin{table*}
\caption{\label{table_TaIrTe}
The coordinates and energy of the Weyl-nodes (W1, W2 and W3) in TaIrTe$_4$ obtained corresponding to Perdew-Burke-Ernzerhof (PBE) exchange-correlation (XC) functionals. The coordinates ($k_x,k_y,k_z$) of the Weyl-nodes W1, W2 and W3 are in the terms of $\left(\frac{2\pi}{a},\frac{2\pi}{b},\frac{2\pi}{c}\right)$.}
\centering
\begin{tabular}{|c|c|c|c|c|c|}
  \hline
  \multicolumn{2}{|c|}{\textbf{W1 point}} & \multicolumn{2}{c|}{\textbf{W2 point}} & \multicolumn{2}{c|}{\textbf{W3 point}} \\
  \hline
  \textbf{($k_x,k_y,k_z$)} & \textbf{Energy (meV)} & \textbf{($k_x,k_y,k_z$)} & \textbf{Energy (meV)} & \textbf{($k_x,k_y,k_z$)} & \textbf{Energy (meV)} \\
  \hline
  ($\pm$0.053, $\pm$0.025, $\pm$0.020) & -46.09 & ($\pm$0.053, $\pm$0.025, 0.000) & -45.78 & ($\pm$0.117, $\pm$0.137, 0.000) & 103.09 \\
  \hline
\end{tabular}
\end{table*}

\begin{figure*}\label{Fig.bands_SrSi2}
\includegraphics[width=0.245\linewidth, height=4.2cm]{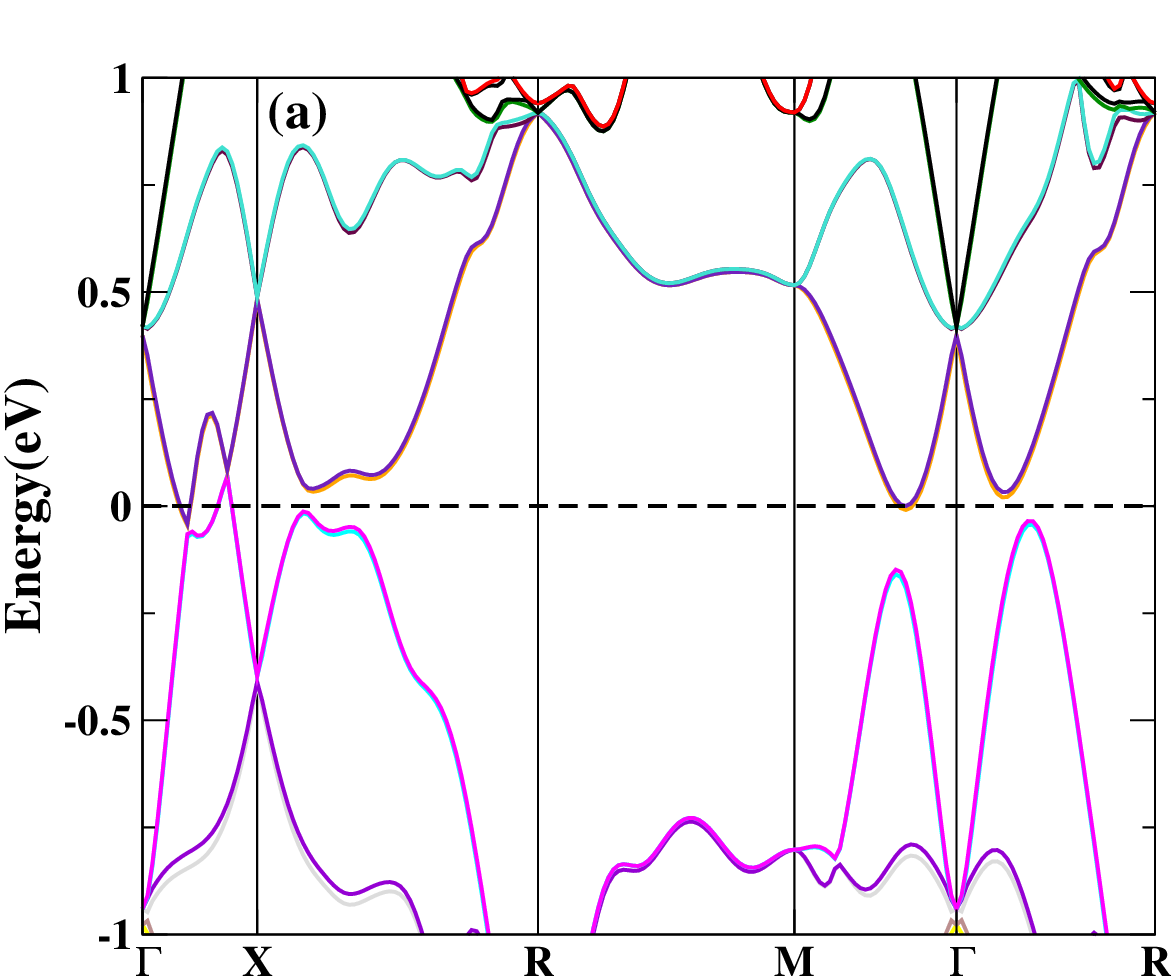}
\includegraphics[width=0.245\linewidth, height=3.95cm]{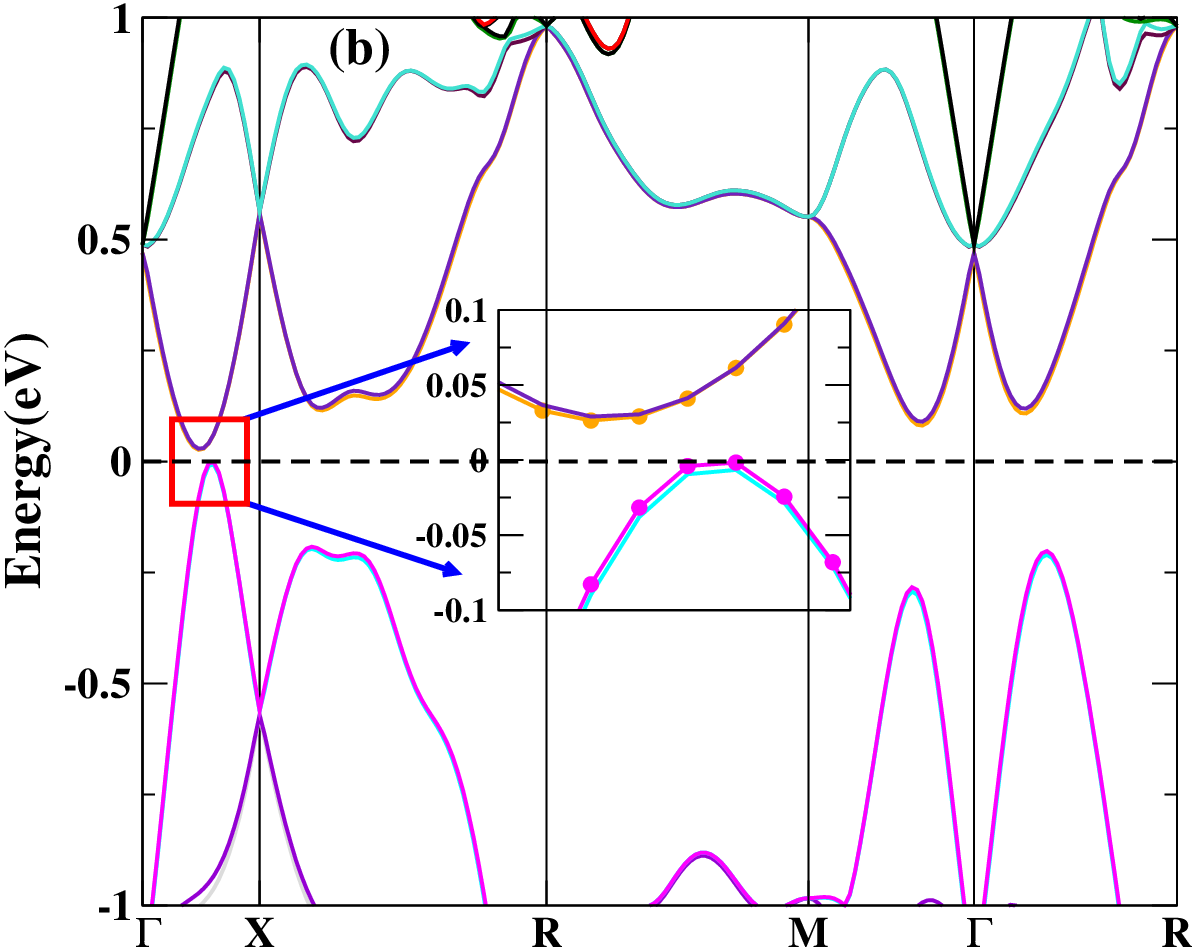}
\includegraphics[width=0.245\linewidth, height=4.2cm]{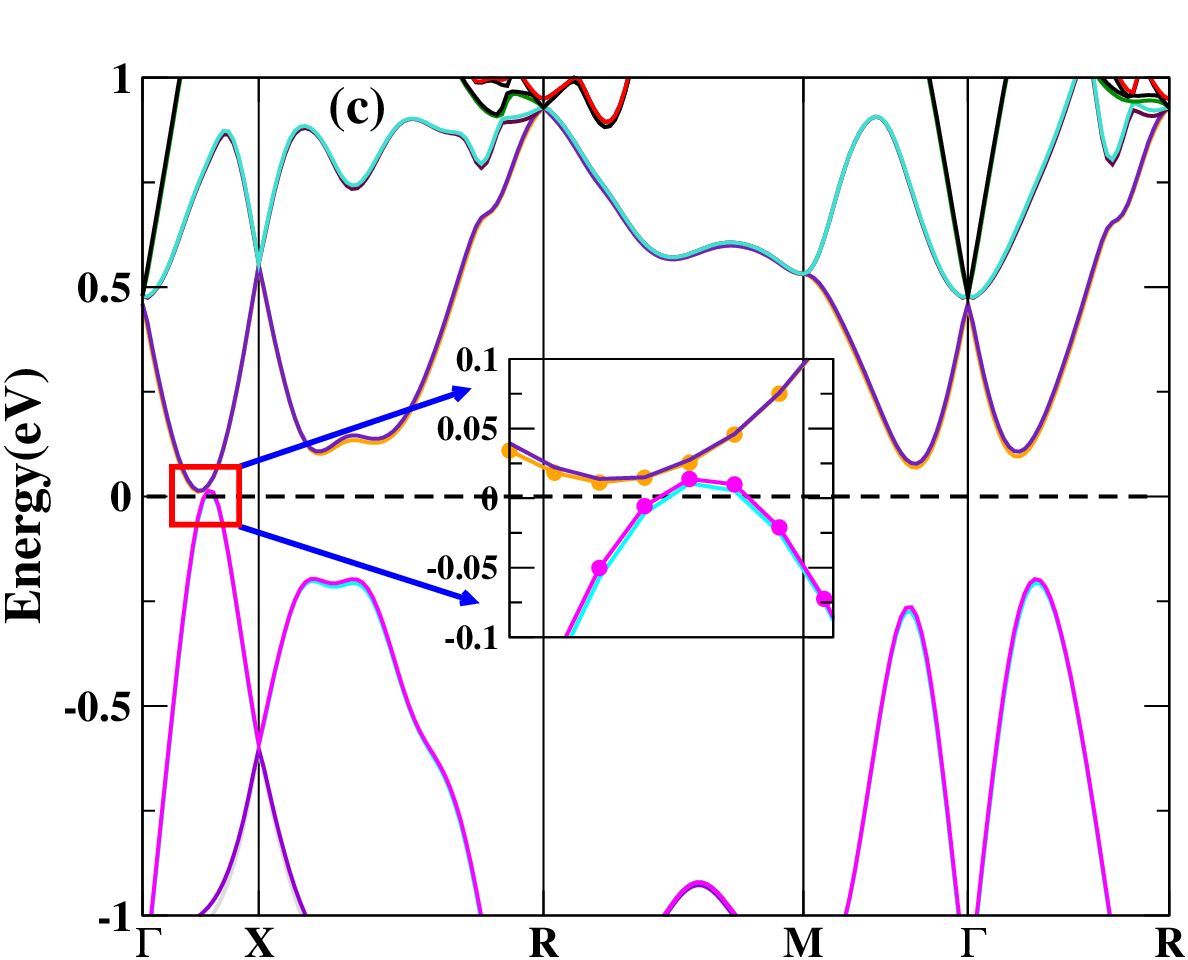}
\includegraphics[width=0.245\linewidth, height=3.95cm]{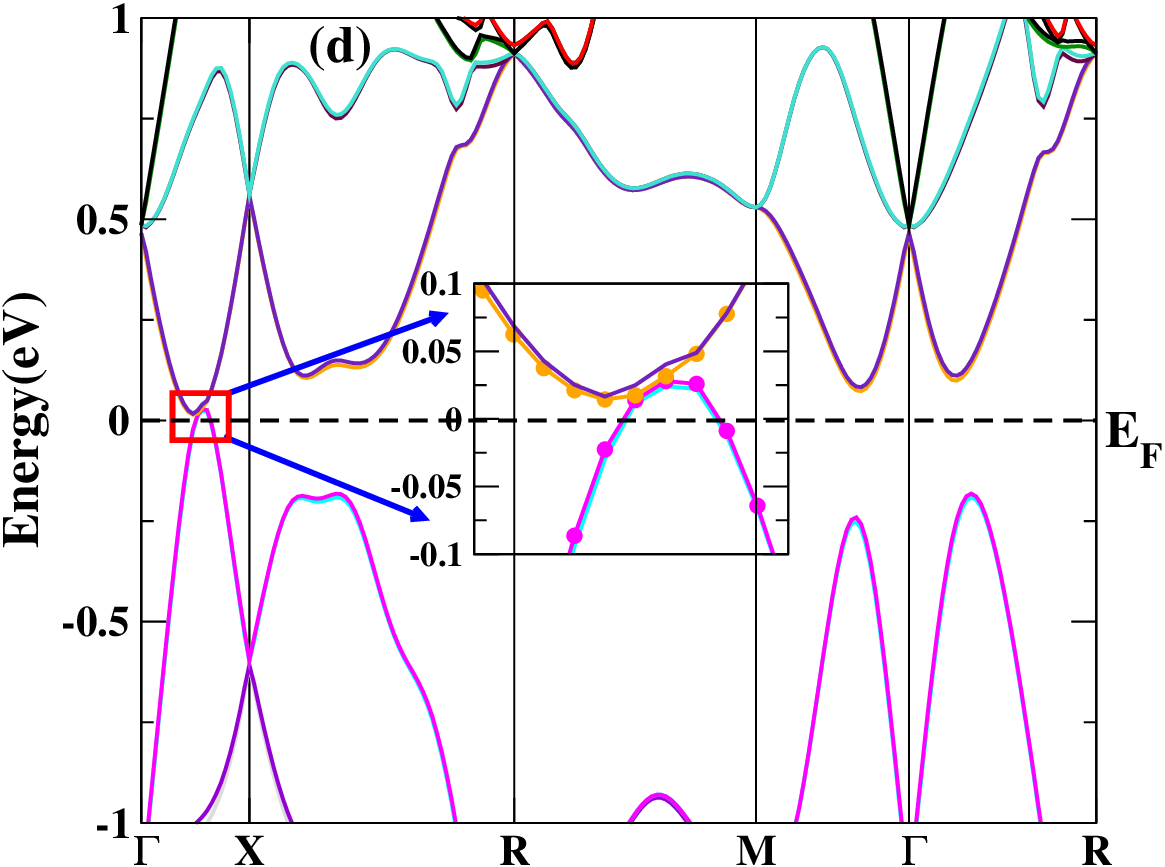}

\caption{\label{Fig.bands_SrSi2}\small{(Color online)
Calculated bulk band structure of SrSi$_2$ with (a) GGA (0\% reduction) (b) TB-mBJ (0\% reduction) and (c) TB-mBJ (1\% reduction) (d) TB-mBJ (1.5\% reduction) functionals in the presence of SOC.
}}
\end{figure*}

\begin{table*}\label{table_SrSi2}
\caption{\label{table_SrSi2}
{The coordinates and energy (in meV) of the Weyl-nodes (W1, W2 and W3) in SrSi$_2$ obtained corresponding to Perdew-Burke-Ernzerhof (PBE) and PBE adapted for solids (PBEsol) exchange-correlation (XC) functionals. The coordinates ($k_x,k_y,k_z$) of the Weyl-nodes W1, W2 and W3 are in the terms of $\left(\frac{2\pi}{a},\frac{2\pi}{b},\frac{2\pi}{c}\right)$. }}
\centering
\begin{tabular}{|p{2.5cm}|c|c|c|c|c|c|}
  \hline
  \multirow{2}{2cm}{\textbf{XC functional}} & \multicolumn{2}{c|}{\textbf{W1 point}} & \multicolumn{2}{c|}{\textbf{W2 point}} & \multicolumn{2}{c|}{\textbf{W3 point}} \\
  \cline{2-7}
  & \textbf{($k_x,k_y,k_z$)} & \textbf{Energy} & \textbf{($k_x,k_y,k_z$)} & \textbf{Energy} & \textbf{($k_x,k_y,k_z$)} & \textbf{Energy} \\
  \hline
       &        ($\pm$0.201, 0.000, 0.000)   & -46.55       & ($\pm$0.343, $\pm$0.061, $\pm$0.061) & 42.76       &    ($\pm$0.367, 0.000, 0.000) & 76.71\\
  PBE  &        (0.000, $\pm$0.201, 0.000)   & -46.55       & ($\pm$0.061, $\pm$0.343, $\pm$0.061) & 42.76       &    (0.000, $\pm$0.367, 0.000) & 76.71\\
       &        (0.000, 0.000, $\pm$0.201)   & -46.55       & ($\pm$0.061, $\pm$0.061, $\pm$0.343) & 42.76       &    (0.000, 0.000, $\pm$0.367) & 76.71\\
  \hline
       &        ($\pm$0.183, 0.000, 0.000)   & -42.78       & ($\pm$0.332, $\pm$0.082, $\pm$0.082) & 43.75       &    ($\pm$0.375, 0.000, 0.000) & 94.87\\
PBEsol &        (0.000, $\pm$0.183, 0.000)   & -42.78       & ($\pm$0.082, $\pm$0.332, $\pm$0.082) & 43.75       &    (0.000, $\pm$0.375, 0.000) & 94.87 \\
       &        (0.000, 0.000, $\pm$0.183)   & -42.78       & ($\pm$0.082, $\pm$0.082, $\pm$0.332) & 43.75       &    (0.000, 0.000, $\pm$0.375) & 94.87 \\
  \hline
\end{tabular}
\end{table*}

{\it Topological nature of SrSi$_2$.|}
SrSi$_2$ belongs to a simple cubic lattice with space group P4$_1$32 (No. 213). The experimental lattice constants of this material are $a = 6.515$ \AA~\cite{pringle1972structure}. 
Imai \textit{et al.}~\cite{imai2005electrical} reported through transport measurements that SrSi$_2$ to be a narrow band gap (10-30 meV) semiconductor, while another experimental study reported it to be a gapless semimetal~\cite{10.1063/1.4772973, 10.3389/fchem.2014.00106}. Recent combined theoretical and experimental studies report that SrSi$_2$ is a narrow-gap semiconductor under ambient conditions.  
In this regard, we first performed electronic structure calculations using the PBE exchange-correlation (XC) functional, employing the optimized lattice constant within PBE ($a$=6.5698 \AA). The corresponding band structure in the presence of SOC is shown in Fig. \ref{Fig.bands_SrSi2} (a). From the figure, it can be seen that within the PBE exchange-correlation (XC) functional, the topmost valence band crosses the Fermi level (E$_F$), indicating that SrSi$_2$ is not a narrow-gap semiconductor. To verify whether a nodal point exists between the topmost valence band and the bottommost conduction band throughout the full Brillouin zone, we searched for nodal points. We have identified three distinct sets of node points, namely W1, W2, and W3, as listed in Table \ref{table_SrSi2}.
It is noted that the node points W1 and W3, obtained through the tight-binding model, were also reported in an earlier study~\cite{singh2018tunable}, whereas the node point W2 was not mentioned in that study.
The newly identified node point W2 forms 12 pairs. To determine whether these 24 node points are indeed Weyl points, we calculated their chiralities ($\chi$) and found that $\chi \in \left\lbrace -1, 1 \right\rbrace$.

As mentioned in the introduction regarding the inconsistencies between two theoretical results~\cite{singh2018tunable, PhysRevB.110.224514} for SrSi$_2$ using the hybrid functional (HSE), we have further investigated whether this band gap truly exists within HSE. 
For this, electronic structure calculations have been performed using the hybrid functional~\cite{PhysRevB.83.235118}, considering the exact exchange parameter $\alpha$=0.25. In this calculation band gap is found to be 71.8 meV. 
Moreover, various studies have shown that hybrid functionals tend to overestimate the experimental band gap and TB-mBJ potential either slightly underestimate or close to experimental band gap~\cite{Band, zimmermann1999electronic, cogan1989optical, enesca2007optical, qiu2011nanowires, kam1982detailed} (For the details see Table IV in the SM). 
Therefore, we need to further examine the band gap using the TB-mBJ potential. 
Using the TB-mBJ potential, we found a gap of $\sim$26 meV between the topmost valence band and the bottommost conduction band (see Fig. \ref{Fig.bands_SrSi2} (b)). It is also important to keep in mind that the GGA functional typically overestimates the experimental lattice parameter, and the case of SrSi$_2$ clearly demonstrates this behavior: the experimental lattice parameter is $a$=6.515 \AA, while the optimized lattice parameter within GGA is $a$=6.5698 \AA.
Therefore, we reduced the optimized lattice parameter by 1\% ($\approx$ 6.5106~\AA) and found that the topmost valence band crosses the Fermi level, indicating metallic behavior (see Fig. \ref{Fig.bands_SrSi2} (c)).
The 1\% reduction in the optimized lattice parameter is nearly the same as the reported experimental value. Hence, it is concluded that at the experimental lattice parameter, SrSi$_2$ is not a narrow gap semiconductor. 
In addition to this, we have also examined the node point at 6.5106 \AA~using TB-mBJ potential. However, within this potential, there is a band gap of 11.14 meV between the topmost valence band and the bottommost conduction band.
However, on reducing lattice parameter by 1.5\% ($\approx$ 6.4712 \AA), we found 12 nodal points with coordinates $\left\lbrace (\pm 0.264, 0, 0)\frac{2\pi}{a}, (0, \pm 0.264, 0)\frac{2\pi}{a}, (0, 0, \pm 0.264)\frac{2\pi}{a} \right\rbrace$ and $\left\lbrace (\pm 0.288, 0, 0)\frac{2\pi}{a}, (0, \pm 0.288, 0)\frac{2\pi}{a}, (0, 0, \pm 0.288)\frac{2\pi}{a} \right\rbrace$, at energies of 18.41 and 31.81 meV, respectively.
Since the node point is obtained with a 1.5\% reduction in the lattice parameter, which is within a reasonable range, the material may be on the verge of becoming a topological semimetal.
Apart from this, we have also optimized the lattice constant using the PBEsol functional. Within this functional, the optimized lattice constant is $a$=6.4977~\AA, which is very close to the experimental lattice parameter. At this lattice parameter ($a$=6.4977~\AA), we have further calculated the node points and found 24 nodes, similar to those obtained with GGA. The corresponding node points are mentioned in Table~\ref{table_SrSi2}.
The above discussion suggests that one should not be surprised if some reports show this material as a topological semimetal while others do not, as variations in sample preparation conditions may lead to small changes in the lattice parameters.
Therefore, our study suggests the careful study  requires for future experimental exploration of this materail.

{\it Verification of topological phase of Cu$_2$XY$_3$.|}
The electronic dispersion curves, both in the absence and presence of SOC, for the Cu$_2$XY$_3$ class of materials are given in Supplementary Material (see Figs. 1 and 2 in SM~\cite{spl}). In the absence of SOC, all the materials in this class exhibit the topmost valence band and the bottommost conduction band appearing to touch each other along X-$\Gamma$ high-symmetric direction. However, in reality, there is a small gap ($\sim$1-8 meV) along X-$\Gamma$ path. 
However, when SOC is included, the degeneracy of the bands near the Fermi level is lifted, resulting in a clear gap between the topmost valence band and the bottommost conduction band in the materials Cu$_2$SiTe$_3$, Cu$_2$GeTe$_3$, Cu$_2$SnSe$_3$, and Cu$_2$SnTe$_3$ along the X-$\Gamma$ path, as evident in the figure.
To verify the topological nature of all materials in this class, we performed calculations to identify the touching points in the full Brillouin zone between the topmost valence band and the bottommost conduction band, both in the absence and presence of SOC. In the absence of SOC, we found that all the materials either host nodal arcs or rings (for details see Table III in SM~\cite{spl}). The single nodal ring obtained for Cu$_2$SnS$_3$ and Cu$_2$SnSe$_3$ from our calculations is consistent with previously reported results~\cite{PhysRevResearch.4.033067}.
In the presence of SOC, the materials Cu$_2$GeSe$_3$, Cu$_2$GeTe$_3$, and Cu$_2$SnSe$_3$ host eight, eight, and four Weyl points, respectively, while Cu$_2$SiTe$_3$ (Cu$_2$GeS$_3$) hosts four small nodal rings (eight nodal-arcs) (for details see Table IV in SM~\cite{spl}). In the case of Cu$_2$SnSe$_3$ and Cu$_2$SnTe$_3$, we did not find any Weyl points or nodal rings. 
However, previously reported results~\cite{PhysRevResearch.4.033067} claim that for Cu$_2$SiTe$_3$, Cu$_2$GeSe$_3$, Cu$_2$GeTe$_3$ and Cu$_2$SnTe$_3$ host eight small nodal rings, sixteen Weyl points, eight Weyl points with four large nodal rings and eight Weyl points, respectively. 
Our results further suggest that predicting topological phases of materials using the TBM constructed from MLWFs may yield inaccurate results.
Therefore, for reliable prediction of topological features, \texttt{PY-NODE} code is better equipped.

{\it Robustness of topological phase against atomic positions.|}
In the above study, we examined the robustness of the topological phases against variations in lattice parameters and exchange-correlation functionals. 
Therefore, it is also interesting to investigate whether the topological phase remains robust against changes in atomic positions that lead to local strain.
Among the studied materials in the Cu$_2$XY$_3$ class, the low atomic number ($Z$) element sulfur (S) is present in Cu$_2$SnS$_3$; hence, the topological phase is expected to be sensitive to atomic positions. To check the sensitivity, we performed electronic structure calculations of Cu$_2$SnS$_3$ with respect to atomic positions in the presence of SOC. 
For this we followed the following strategies: i) Applied local strain to the site of one atom at a time while keeping the other atoms relaxed; ii) Applied local strain to the sites of two atoms at a time while keeping the remaining two atoms relaxed; iii) Applied local strain to the sites of three atoms at a time while keeping one atom relaxed; iv) Applied local strain to the site of each atom individually. 
Throughout this process, the atomic positions of the Cu, Sn, S1, and S2 atoms are varied by $\pm$0.01 relative to their respective relaxed positions. 
Within a variation of $\pm$0.01, and considering all such combinations of atoms following the above-mentioned strategies, the Weyl phase is found to be robust.  
However, by considering the atomic positions ($\mathrm{Cu}_x^R$, 0.5, $\mathrm{Cu}_z^R$), (0, 0, $\mathrm{Sn}_z^R - 0.019$), ($\mathrm{S1}_x^R + 0.01$, 0.0, $\mathrm{S1}_z^R - 0.015$), and (0.5, 0, $\mathrm{S2}_z^R + 0.03$), we found the nodal arc phase (here, \(\mathrm{Cu}_i^R\), \(\mathrm{Sn}_i^R\), and \(\mathrm{S1(2)}_i^R\), with \(i = x,y,z\), are the relaxed positions). 
The bond distance between Cu and S1 atoms is 2.355~\AA, and the bond angle S1--Cu--S2 is 107.89$^\circ$ for the above configuration. However, for the relaxed structure, the bond distance and bond angle are 2.314~\AA{} and 108.13$^\circ$, respectively. 
This suggests that even a small change in bond distance and bond angle from the relaxed structure driving the Weyl phase to nodal arc phase indicates that the topological phase is sensitive to atomic positions.
In addition to this, we have found the nodal arc phase at other atomic positions also (see Table V in the SM~\cite{spl} for details).


{\it Conclusions.|}
Experimental efforts to verify the topological phases of materials predicted by theoretical catalogs have been made, but not all have yielded positive results.
Therefore, in this study, we have attempted to identify possible reasons for the discrepancies between experimental and theoretical results, as well as between different theoretical results, using various examples. Our results highlight the limitations of deriving results from MLWF-based tight-binding models, as demonstrated through the examples of TaIrTe$_4$, SrSi$_2$, and Cu$_2$XY$_3$ (X = Si, Ge, Sn; Y = S, Se, Te). Furthermore, the sensitivity of the topological phase to exchange-correlation (XC) functionals and lattice parameters is also demonstrated. It is found that the topological phase of SrSi$_2$ is highly sensitive to both XC functionals and lattice parameters.
In addition to this, we have also studied the sensitivity of the topological phase with respect to atomic positions. 
It has been observed that a small change in atomic positions from the relaxed structure leads the system from a Weyl phase to a nodal arc phase, indicating that the topological phase is highly sensitive to atomic positions in Cu$_2$SnS$_3$.
Our study provides direction for the credible prediction of the topological phases of materials by studying the robustness of these phases against variations in lattice parameters, atomic positions, and exchange-correlation functionals.

\bibliographystyle{apsrev4-2}
\bibliography{ref} 

\end{document}